\documentstyle[prd,aps,epsfig,eqsecnum,amssymb,amsmath,floats]{revtex}
\begin{document}
\title{ 
Constraints on diffuse neutrino background from primordial 
black holes
}
\author{
E. V. Bugaev and K. V. Konishchev 
}
\address{
Institute for Nuclear Research, Russian Academy of Sciences, 
Moscow 117312, Russia 
}
\date{
\today
}
\maketitle

\begin{abstract}
We calculated the energy spectra and the fluxes of electron neutrinos emitted
in the process of the evaporation of primordial black holes (PBHs) in the early
universe. It was assumed that PBHs are formed by a blue power-law spectrum
of primordial density fluctuations. We obtained the bounds on the spectral index
of density fluctuations assuming validity of the standard picture of 
gravitational collapse and using the available data of several experiments 
with atmospheric and solar neutrinos. The comparison of our results with the
previous constraints (which had been obtained using diffuse photon background 
data) shows that such bounds are quite sensitive to an assumed form of the 
initial PBH mass function.
\end{abstract}

\section{Introduction}
\label{sec:int}
Some recent inflation models (e.g., the hybrid inflationary scenario \cite{1})
predict the "blue" power-spectrum of primordial density fluctuations. In turn,
as is well known, the significant abundance of primordial black holes (PBHs) 
is possible just in the case when the density fluctuations have an $n>1$ 
spectrum ($n$ is the spectral index of the initial density fluctuations, 
$n>1$ spectrum is, by definition, the "blue perturbation spectrum").
 
   Particle emission from PBHs due to the evaporation process predicted by
Hawking \cite{2} may lead to observable effects. Up to now, PBHs have not been 
detected, so the observations have set limits on the initial PBH abundance
or on characteristics of a spectrum of the primordial density fluctuations.
In particular, PBH evaporations contribute to the extragalactic neutrino 
background. The constraints on an intensity of this background (and, 
correspondingly, on an PBH abundance) can be obtained from the existing experiments
with atmospheric and solar neutrinos. The obtaining of such constraints is
a main task of the present paper.

 The spectrum and the intensity of the evaporated neutrinos depend heavily on 
the PBH's mass. Therefore, the great attention should be paid to the calculation
of the initial mass spectrum of PBHs. We use in this paper the following 
assumptions leading to a prediction of the PBH's mass spectrum.

1. The formation of PBHs begins only after an inflation phase when the universe
returns to the ordinary radiation-dominated era. The reheating process is such
that an equation of state of the universe changes almost instantaneously 
into the radiation type (e.g., due to the parametric resonance \cite{3}) after the 
inflation.

2. It is assumed, in accordance with analytic calculations \cite{4,5} that a critical 
size of the density contrast needed for the PBH formation, $\delta_c$, is 
about $1/3$. Further, it is assumed that all PBHs have mass roughly equal
to the horizon mass at a moment of the formation, independently of the 
perturbation size.

3. Summation over all epochs of the PBH formation can be done using the
Press-Schechter formalism \cite{6}. This formalism is widely used in the standard
hierarchial model of the structure formation for calculations of the mass
distribution functions (see, e.g., \cite{7}).

It was shown recently that near the threshold of a black hole formation the gravitational collapse
behaves as a critical phenomenon \cite{8}. In this case the initial mass function will be quite 
different from that follows from the standard calculations of refs.~\cite{5,7}.The first calculations
of the PBH initial mass function (e.g., in ref.\cite{8}) have been done under the assumption that 
all PBHs form at the same horizon mass (by other words, that all PBHs form at the smallest horizon 
scale immediatly after reheating). The initial PBH mass spectrum for the case of the critical
collapse, based on the Press-Schechter formalism, was obtained in the recent work \cite{9}.

The calculations in the present paper are based on the standard \cite{4,5} picture of the 
gravitational collapse leading to a PBH formation. The case of the critical collapse will
be considered in a separate work.

The plan of the paper is as follows.

In Sec.\ref{sec:PBHsp} we give, for completeness, the brief derivation of a general formula
for the initial PBH mass spectrum. The final expression is presented in a form, which is
valid for an arbitrary relation between three physical values:
the initial PBH mass, the fluctuation mass (i.e., the mass of the perturbed region) at a moment of 
the collapse, and the density contrast in the perturbed region (also at a moment of the collapse).
This expression contains the corresponding results obtained in refs.~\cite{7,9} as particular cases.
As in refs.~\cite{7,9}, the derivation is based on the linear perturbation theory and on the assumption 
that a power spectrum of the primordial fluctuations can be described by a power law.

In Sec.\ref{sec:neutbackgr} we derive the approximate formula for a calculation of the extragalactic
neutrino background from PBH evaporations. We do not use cosmological models of the inflation and of the spectrum
of primordial fluctuations, so there are two free parameters: the reheating temperature and the spectral index.
At the end of the section, some examples of instantaneous neutrino spectra from the evaporation of an 
individual black hole are presented.

In Sec.\ref{sec:constr} we determine the explicit dependence of an background intensity on the spectral
index (by normalization of standard deviation of the density contrast at horizon crossing on COBE data). 
After this, numerical calculations of the neutrino background become possible. For background neutrino energies
$\sim 10-100\;MeV$ we studied relative contributions to the background intensity of different cosmological
redshifts (and it is shown that at high reheating temperatures the characteristic values of the redshifts are very large).
Further, several results of numerical calculations of the neutrino background spectra are presented.
Possibilities of constraining the spectral index using experiments with neutrinos of natural origin are discussed.

In Sec.\ref{sec:res} the spectral index constraints, followed from our calculations and from available data
of the neutrino experiments, are given. It is proved that, almost in all intervals of the reheating
temperatures considered in this work, effects of an background neutrino absorption in the space are small and can be neglected.

\section{The initial mass spectrum of PBHs}
\label{sec:PBHsp}
As it pointed out in the {\bf Introduction}, we use the Press-Schechter formalism which allows to carry out the 
summation over all epochs of PBHs formation. According to this formalism,
the mass spectrum of density fluctuations ( i.e., the number density of regions with mass 
between $M$ and $M+dM$ ) is calculated by the formula
\begin{equation}
\label{1}
n\left( M,\delta _c \right)dM=\frac{\rho _i }{M}
\left| \frac{\partial \beta}{\partial M}(M,\delta _c )\right| dM\;\;\;.
\end{equation}
Here, $\beta (M,\delta _c)$ is the fraction of regions having sizes 
larger than $R$ and density contrast larger than $\delta _c$ ,
\begin{equation}
\label{2}
\beta(M,\delta _c)=2\int^\infty_{\delta _c} P(M,\delta)d\delta\;\;\;,
\end{equation}
\begin{equation}
\label{3}
P(M,\delta)=\frac{1}{\sqrt{2\pi}\sigma _R (M)}\exp\left(-\frac{\delta^2}
{2\sigma _R^2 (M)} \right)\;\;\;.
\end{equation}
Here, $\delta$ is the initial density contrast, $\sigma_R$ is the standard
deviation of the density contrast of the regions with size $R$ and mass $M$.

It is convenient to introduce the double differential distribution 
$n(M,\delta)$, the integral over which gives the total number of fluctuated
regions,
\begin{equation}
n=\int\!\!\!\!\int n(M,\delta)dMd\delta \;\;\;\;,\;\;\;
n(M,\delta_c)=\int n(M,\delta)d\delta\;\;\;.
\end{equation}
Using Eqs.(\ref{1}-\ref{3}), for $n(M,\delta)$ one has the expression
\begin{equation}
\label{25}
n(M,\delta)=\sqrt{\frac{2}{\pi}}\frac{\rho_i}{M}\frac{1}{\sigma_R^2 (M)}\frac{\partial \sigma_R}{\partial M}
\left|\left(\frac{\delta^2}{\sigma_R^2 (M)}-1\right)\right|\exp\left(
-\frac{\delta^2}{2\sigma_R^2 (M)}\right)\;\;\;.
\end{equation}

To obtain the mass spectrum of PBHs one must introduce the $M_{BH}$ variable.
Besides, we will use the variable $\delta '$, connected with $\delta$
by the relation
\begin{equation}
\delta ' = \delta \left( \frac{M}{M_i} \right) ^{2/3}\;\;.
\end{equation}
Here, $M_i$ is the horizon mass at the moment of a beginning of the growth of density fluctuations.
This new variable is the density contrast at the moment of the collapse.
The new distribution function is 
\begin{equation}
\label{*}
n_{BH}(M_{BH},\delta ') = n(M_{BH},\delta ) \frac{d\delta}{d\delta '}
\frac{d M}{d M_{BH}}\;\;.
\end{equation}
Further, we assume that there is some functional connection between $M_{BH}$, $M$ and $\delta '$:
\begin{equation}
\label{0}
M_{BH}=f(M,\delta ')\;\;.
\end{equation}
In this case one can rewrite Eq.(\ref{*}) in the form:
\begin{equation}
\label{29}
n_{BH} (M_{BH},\delta ') = n\left(M,\delta ' \left(\frac{M}{M_i}\right)^{-2/3}\right)\cdot \left(\frac{M}{M_i} \right)^{-2/3}
\cdot \frac{1}{{d f(M,\delta ')}/{d M}} 
\end{equation}
and the PBH mass spectrum is given by the integral
\begin{equation}
\label{210}
n_{BH}(M_{BH}) = \int n_{BH}(M_{BH},\delta ') d\delta ' \;. 
\end{equation}
Now, to connect the mass spectrum with the spectral index one can use the relations 
\begin{equation}
\label{**}
\sigma_R = \sigma_H (M) \left(\frac{M}{M_i} \right)^{-2/3}\;\; ; \;\; \sigma_H (M) \sim M^{\frac{1-n}{6}}
\end{equation}
(see Sec.\ref{sec:constr} for details).
Using Eqs.(\ref{**}) in the expression (\ref{25}) for $n(M,\delta)$ one obtains:
\begin{equation}
n(M,\delta) = -\frac{n+3}{6}\cdot \sqrt{\frac{2}{\pi}} \frac{\rho_i}{M_i^2} \frac{1}{\sigma_H}
\left(\frac{M}{M_i} \right)^{-4/3} \left|\frac{\delta^2}{\sigma_R^2} -1\right| e^{-{\delta^2}/{2\sigma_R^2}}\;\;.
\end{equation}
Substituting this expression in the r.h.s. of the Eq.(\ref{29}) and using Eq(\ref{210}) one has: 
\begin{equation}
\label{213}
n_{BH}(M_{BH})= 
\frac{n+3}{6}\cdot\sqrt{\frac{2}{\pi}}\rho_i \int \frac{1}{M^2 \sigma_H} \left|\frac{{\delta '} ^2}{\sigma_H^2}-1 \right|
e^{-\frac{{\delta '} ^2}{2\sigma_H^2}} \frac{1}{d f(M,\delta ')/ d M} d{\delta '} \;\;.
\end{equation}
Values of $M$ in r.h.s of Eq.(\ref{213}) are expressed through $M_{BH}$, $\delta '$ by the relation (\ref{0}).

Formula (\ref{213}) is the final expression for the PBH mass spectrum, the main result of this Section. It is valid for any relation 
between PBH mass $M_{BH}$, mass of the original overdense region $M$ and the density contrast $\delta '$.   

In the particular case of a near critical collapse 
\begin{equation}
\label{214}
f(M, \delta ') = M_i^{1/3}M^{2/3} k (\delta ' -\delta_c )^{\gamma_k}\equiv \xi M^{1/3}_i M^{2/3}
\end{equation}
and
\begin{equation}
M=M_{BH}^{3/2}M_i^{-1/2}\xi^{-3/2}\;\; ; \;\; d f/d M =\frac{2}{3} M_i^{1/2}M_{BH}^{-1/2}\xi^{3/2}\;\;.
\end{equation}

Eq.(\ref{214}) can be rewritten in the form, derived in the ref.~\cite{8}, using the connection between $M$ and the horizon mass
$M_{h}$ (which is equal to a fluctuation mass at the moment when the fluctuation crosses horizon):
\begin{equation}
\label{eee}
M_h=M_i^{1/3}M^{2/3}\;\;.
\end{equation}

The resulting formula for the spectrum,
\begin{equation}
\label{eq}
n(M_{BH}) =
\frac{n+3}{4} \sqrt{\frac{2}{\pi}} \sqrt{M_i} M_{BH}^{-5/2} \int^1_{\delta_c} \frac{1}{\sigma_H} \left| \frac{\delta '{}^2}
{\sigma_H^2}-1 \right| e^{-\frac{\delta^2}{2\sigma_H^2}}\xi^{3/2}d \delta ' \;\;,
\end{equation}
coincides, as can be easily proved, with the expression derived in ref.~\cite{9}.

The mass spectrum formula for the  Carr-Hawking collapse can be obtained analogously, with the relation 
$f(M,\delta ')= \gamma^{1/2}M_i^{1/3}M^{2/3}$, or from Eq.(\ref{eq}) using the substitutions
\begin{equation}
\gamma_k\to 0\;\;,\;\;k\to \gamma^{1/2}\;\;,\;\;\delta_c\to\gamma
\end{equation}
and the approximate relation
\begin{equation}
\int^1_{\gamma} d\delta ' \left(\frac{\delta '{}^2}{\sigma_H^2} -1 \right) e^{-\frac{\delta '{}^2}{2\sigma_H^2}}
\approx \gamma e^{-\frac{\gamma^2}{2\sigma_H^2}}\;\;.
\end{equation}
In this case one obtains the expression obtained in ref.~\cite{7}:
\begin{equation}
\label{220}
n_{BH}(M_{BH})=\frac{n+3}{4}\sqrt{\frac{2}{\pi}}\gamma^{7/4}\rho_iM_i^{1/2}
M_{BH}^{-5/2}\sigma_H^{-1}\exp\left(-\frac{\gamma^2}{2\sigma_H^2}\right)\;.
\end{equation}
One can see from Eqs.(\ref{eq}) and (\ref{220}) that the PBH mass spectrum following from
the Press-Shechter formalism has quasi power form in both considered 
cases $\left(\sim M_{BH}
^{-5/2}\right)$. In Carr-Hawking case $M_{BH}^{min}\sim M_h$ , in contrast 
with this in the critical collapse case it is possible that $M_{BH}\ll M_h$ . 
However , the low mass part of the PBH spectrum is suppressed by the factor 
$\xi^{3/2}$ in the integral in Eq. (\ref{eq}).

\section{Neutrino diffuse background from PBHs}
\label{sec:neutbackgr}
The starting formula for a calculation of the cosmological background 
from PBH evaporations is \cite{10}
\begin{equation}
\label{31}
S(E)=\int n_{com}\frac{1}{4\pi a_0^2\rho^2_{\mathstrut}}f\left(E(1+z)
\right)dV_{com}\;\;\;.
\end{equation}
Here , $n_{com}$ is the comoving number density of the sources (in our case
the source is an evaporating PBH of the definite mass $m$ ), $a_0$ is the 
scale factor at present time, $t=t_0$, $f(E)$ is a differential energy 
spectrum of the source radiation, $V_{com}$ is a comoving volume of the
space filled by sources, therefore 
\begin{equation}
\label{32}
dV_{com}=a_0^3\frac{\rho^2 d\rho}{\sqrt{1-k\rho^2}}d\Omega\;\;\;.
\end{equation} 
Here, $k$ is the curvature coefficient,
and $\rho$ is the radial comoving coordinate. Using the change of the variable,
\begin{equation}
\label{33}
\frac{d\rho}{\sqrt{1-k\rho^2}}=\frac{dt}{a}\;\;\;,
\end{equation}
the comoving number density can be expressed via the initial density $n_i$,
\begin{equation}
\label{34}
n_{com}=n_{phys}(t_0)=n_i\left(\frac{a_i}{a}\right)^3\left(\frac{a}{a_0}\right)
^3=n_i\left(\frac{a_i}{a_0}\right)^3\;\;\;\;.
\end{equation}
Substituting Eqs. (\ref{32}) - (\ref{34}) in Eq. (\ref{31}) one obtains
\begin{equation}
S(E)=n_i\int dt\frac{a_0}{a}\left(\frac{a_i}{a_0}\right)^3 f\left(
E(1+z)\right)\;\;\;.
\end{equation}
In our concrete case the source of the radiation is a Hawking evaporation:
\begin{equation}
\label{36}
n_i f\left(E(1+z)\right) = \int dm\, n_{BH}(m,t)f_H\left(E(1+z),m\right)\;\;\;.
\end{equation}
Here , $n_{BH}(m,t)$ is the PBH mass spectrum at any moment of time,
$f_H(E,m)$ is the Hawking function \cite{2},
\begin{equation}
\label{37}
f_H\left(E,m\right)=\frac{1}{2\pi\hbar}\frac{\Gamma_s(E,m)}{exp\left(\frac{8\pi
GEm}{\hbar c^3}\right)-(-1)^{2s}}\;\;\;.
\end{equation}
Here, $\Gamma_s(E,m)$ is the coefficient of the absorption by a black hole
of a mass $m$, for an particle having spin s and energy $E$.

  Initial spectrum of PBHs is given by Eq. (\ref{220}). The minimum value of PBH mass
is $\gamma^{1/2}_{\mathstrut}M_i^{\mathstrut}$, so we must add
to the initial spectrum expression the step factor $\Theta(m_{BH}^{\mathstrut}-
\gamma^{1/2}_{\mathstrut}M_{i}^{\mathstrut})$ . The connection of the initial 
mass value $M_{BH}$ and the value at any moment $t$ is determined by the solution 
of the equation \cite{11}:
\begin{equation}
\label{38}
\frac{d m}{d t}=-\frac{\alpha (m)}{m^2}\;\;.
\end{equation}
The function $\alpha (m)$ accounts for the degrees of freedom of evaporated particles
and determines the lifetime of a black hole. In the approximation $\alpha = const$
the solution of Eq.(\ref{38}) is: 
\begin{equation}
\label{310}
M_{BH}\cong\left(3\alpha t+m^3\right)^{1/3}\;\;\;.
\end{equation}
This decrease of PBH mass leads to
the corresponding evolution of a form of the PBH mass spectrum. At any moment one has
\begin{equation}
\label{311}
n_{BH}(m,t)dm=\frac{m^2}{\left(3\alpha t+m^3\right)^{2/3}}n_{BH}\left((3\alpha
t+m^3)^{1/3}\right)\times
\Theta\left[m-\left((\gamma^{1/2}M_{i})^3-3\alpha t\right)^{1/3}\right]dm.
\end{equation}
Substituting Eqs. (\ref{37}), (\ref{311}) in the integral in Eq.(\ref{36}) , we obtain the
final expression for the spectrum of the background radiation:
\begin{eqnarray}
\label{28}
S(E)=\frac{c}{4\pi}\int dt \frac{a_0}{a}\left(\frac{a_i}{a_0}\right)^3
\int dm \frac{m^2}{(3\alpha t+m^3)^{1/3}}n_{BH}\left[\left(3\alpha t+m^3\right)
^{1/3}\right]\times\nonumber\\
\\
\Theta\left[(m-\left((\gamma^{1/2}M_i)^3-3\alpha t\right)^{1/3}\right]
f_H(E(1+z),m)\;\;\;.\nonumber
\end{eqnarray} 

One should note that the corresponding expressions for the spectrum in
refs.~\cite{12,13} contain the factor $\left(\frac{a_i}{a}\right)^3$ instead of the
correct factor $\left(\frac{a_i}{a_0}\right)^3$. It leads to a strong 
overestimation of large $z$ contributions in $S(E)$ (see below, Fig.3).

It is convenient to use in Eq.~(\ref{28}) the variable $z$ instead of $t$. 
In our case ($\Omega_{\Lambda}=\Omega_{K}=0$) we have
\begin{eqnarray}
\frac{dt}{dz}=-\frac{1}{H_0(1+z)}\left(\Omega_{m_0}(1+z)^3+\Omega_{r_0}(1+z)^4
\right)^{-1/2}\;\;\;,\\
\nonumber\\
\Omega_{r_0}=(2.25\cdot 10^4h^2)^{-1}\;\;\;\;,\;\;\;\Omega_{m_0}=1-
\Omega_{r_0}\;\;\;.\nonumber
\end{eqnarray}
The factor $(\frac{a_i}{a_0})^3$ can be expressed through the value of $t_{eq}$:
\begin{equation}
\left(\frac{a_i}{a_0}\right)^3\simeq(1+z_{eq})^{-3}\left(\frac{t_i}{t_{eq}}
\right)^{3/2}\simeq H_0^{-3/2}\left(2.25\cdot10^4h^2\right)^{-3/4}t_i^{3/2}\;\;\;.
\end{equation}
Integrating over PBH's mass in Eq.~(\ref{28}) one obtains finally, after the change of the variable $t$ on $z$, the integral
over~$z$:
\begin{equation}
\label{313}
S(E)=\int d\log_{10}(z+1)\,F(E,z)\;\;\;.
\end{equation}

In analogous calculations of the photon diffuse background integral over $z$ in the expression
for $S(E)$ is cut off at $z=z_0\approx 700$ because for larger $z$ the photon
optical depth will be larger than unity \cite{14}. In contrast with this , interactions
of neutrinos with the matter can be neglected up to very high values of $z$.
Therefore the neutrino diffuse background from PBH evaporations is much more
abundant.
 The neutrino absorption effects are estimated below, in Sec.\ref{sec:res}.

The evaporation process of a black hole with 
not too small initial mass is almost an explosion. So, for a calculation of spectra of evaporated particles 
with acceptable accuracy it is enough to know the value of $\alpha$ for an initial
value of the PBH mass only. Taking this into account and having in mind the steepness of the PBH mass spectrum, 
we use the approximation
\begin{equation}
\alpha (m)=\alpha (M_{BH}^{min} )=\alpha(\gamma^{1/2} M_i)\;\;,
\end{equation}
and just this value of $\alpha $ is meant in the expressions (\ref{310})-(\ref{28}).

The very detailed calculation of the function $\alpha (m)$ was carried out
in the works \cite{15,16}. Here we use the simplified approach in which $\alpha (m)$
is represented by the dependence
\begin{equation}
\label{36a}
\alpha=\alpha_0 + \sum\limits_i a_i\cdot 10^{25}\cdot \Theta(b_i-log_{10}(m)).
\end{equation}
\begin{table}[htb] 
\caption{Coefficients $a_i$ and $b_i$ in Eq.(\ref{36a})  \label{tab:t1}}
\begin{tabular}{@{\hspace{.4in}}lllllllllll@{\hspace{.4in}}}

$i$   & $\mu$     & u,d       & s      & g      & $\tau$      & c      &  b      & $W^{\pm}$       & $Z^0$ & t       \\
\hline
$a_i$ & 3.12      &  9.36     & 9.36   & 4.96   &  3.12       & 9.36   &  9.36   &  1.86           & 0.93    & 9.36  \\
$b_i$ & 14.47     &  14.01    & 13.77  & 13.67  &  13.23      & 13.20  &  12.79  &   11.57         & 11.55   & 11.27 \\ 

\end{tabular}
\end{table} 

Here, $\Theta (x)$ is the Heaviside step function. Coefficient $\alpha_0$ gives the summary
contribution of $e^{+}$, $e^{-}$, $\nu$ and $\gamma$ and is equal to $8.42\cdot 10^{25}\; g^3 sec^{-1}$
\cite{11,17}. All other coefficients are collected in the Table~\ref{tab:t1}.

The coefficients $b_i$ determine
the value of the PBH mass beginning from which particles of $i$-type can be evaporated. This value, $M_{BH}^b$,
is obtained from the relations 
\begin{equation}
\frac{10^{13}}{M_{BH}^b (g)}\cong T_{BH}^b (GeV) \cong \frac{m_i}{3}\;.
\end{equation}
So, 
\begin{equation}
b_i =log_{10} M_{BH}^{b}\;\;.
\end{equation}
\begin{figure}
\caption{The parameter $\alpha$ in the (\ref{36a}) for the rate of the PBH mass loss, as a function of the PBH mass. Solid line is the present result, dashed line is drawn using Eq.(7) of Ref.~[16].}
\label{fig:fig1}
\epsfig{file=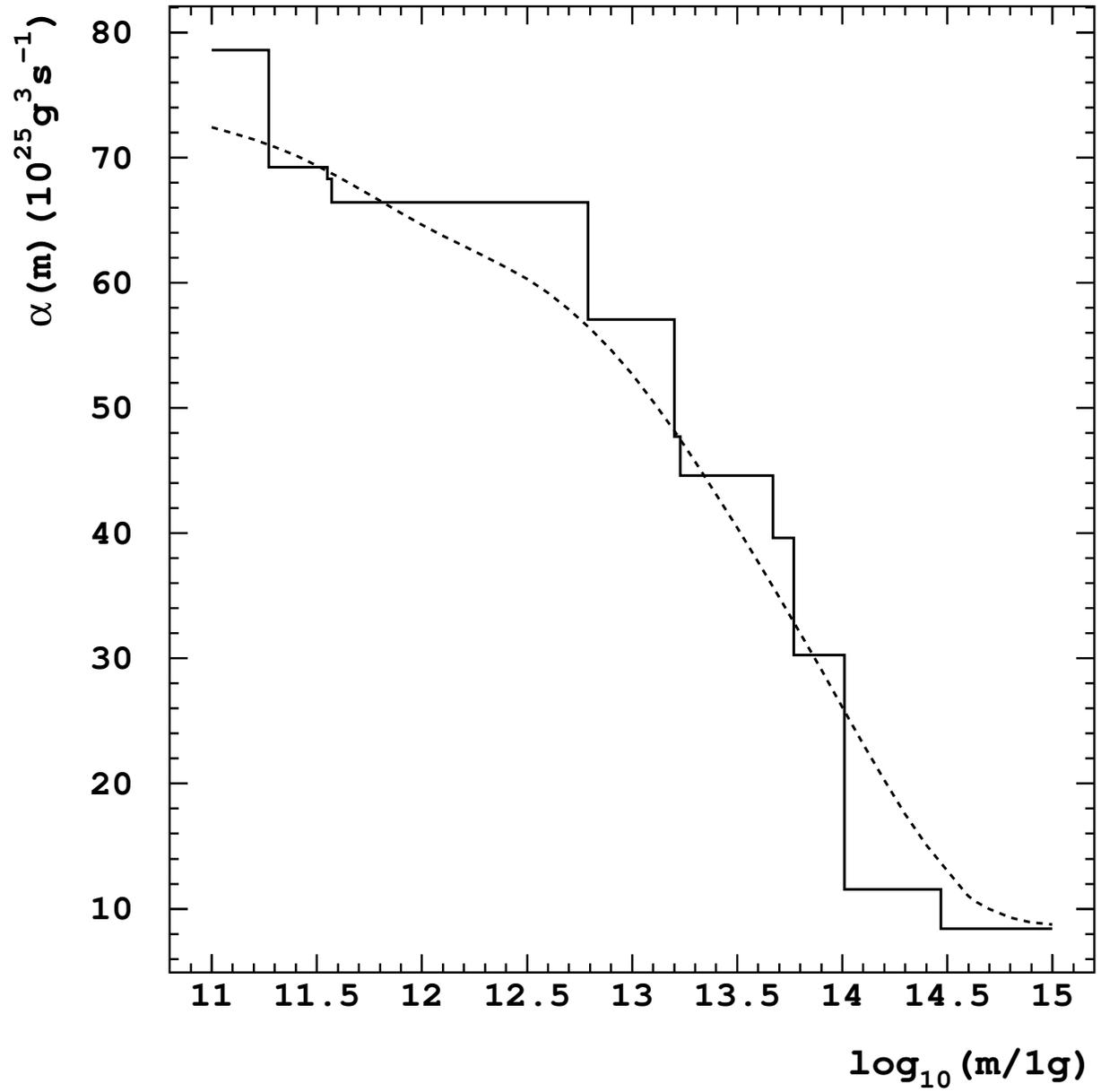,width=\columnwidth}
\end{figure}

The resulting function $\alpha (m)$ is shown on Fig.\ref{fig:fig1} For comparison, the corresponding dependence
from the work \cite{16} (drawn using the Eq.7 of \cite{16}) is also shown.

The formula (\ref{28}), as it stands, takes into account the contribution to the neutrino background solely from
a direct process of the neutrino evaporation. If we suppose that particles evaporated by a black hole 
propagate freely, the calculation of other contributions to the neutrino background can be performed using our
knowledge of particle physics \cite{15,16}.

\begin{figure}
\caption{Instantaneous electron neutrino spectra from a PBH evaporation, for the several channels of the neutrino production.}
\label{fig:fig2}
\epsfig{file=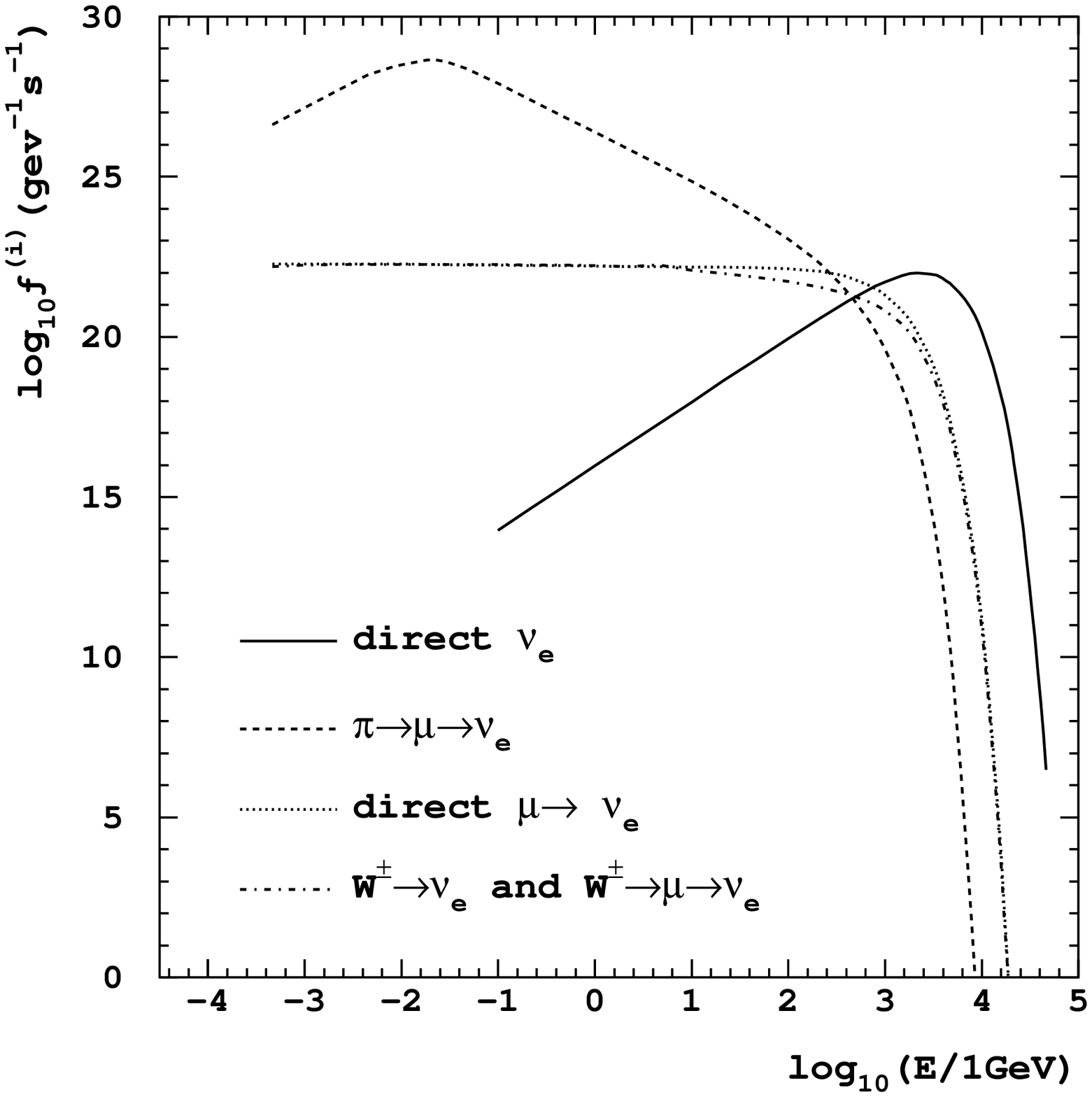,width=\columnwidth}
\end{figure}

On Fig.\ref{fig:fig2} the typical result of our calculation of instantaneous neutrino spectra from evaporating black hole is shown.
The spectrum of the straightforward (direct) $(\nu_e+\tilde\nu_e)$ - emission is given by the Hawking function $f_H (E,m)$, Eq.(\ref{37}).
The $(\nu_e + \tilde\nu_e)$ spectrum arising from decays of $(\mu^{+} + \mu^{-})$ evaporated directly is calculated by the formula
\begin{equation}
f^{(\mu)}(E,m) = \int f_H (E_{\mu},m)\frac{d n_{\nu} (E_{\mu},E)}{d E}d E_{\mu}\;\;,
\end{equation} 
where $d n^{\mu}_{\nu}/d E$ is the neutrino spectrum in a $\mu$ - decay. To evaluate the electron neutrino spectrum resulted from fragmentations of 
evaporated quarks, we used the simplest chain:
\begin{equation}
(u,d) - \mbox{quarks} \to \pi \to \mu \to \nu_e\;\;,
\end{equation}
and the formula 
\begin{equation}
f^{(q)} (E,m)= \int f_H (E_q,m)\frac{d n^q_{\pi}(\xi)}{d \xi}\cdot\frac{d \xi}{d E_q}\cdot \frac{d n_{\mu}^{\pi}(E_{\mu},E_{\pi})}{d E_{\mu}}
\cdot\frac{d n^{\mu}_{\nu}(E_{\mu},E)}{d E} d E_{q} d E_{\pi} d E_{\mu} \;\;.
\end{equation}
Here, ${d n^q_{\pi}}/{d \xi}$ is the $q\to \pi$ fragmentation function, for which the simple parametrization was taken:
\begin{equation}
\frac{d n^q_{\pi}}{d \xi} =\frac{15}{16}(\xi - 1)^2 \xi^{-3/2}\;\;,\;\; \xi=\frac{E_{\pi}}{E_q}\;\;,
\end{equation} 
and ${d n_{\mu}^{\pi}}/{d E_{\mu}}$ is the $\mu$ - spectrum in
a decay $\pi \to \mu + \nu_{\mu} $.

Analogous formulas are used for calculations of the neutrino spectrum from other channels of the neutrino production, 
for instance from  the decays of evaporated $W$-bosons
($W\to e + \nu_e\;\;,$ $W\to \mu \to \nu_e$).

The relative contribution of different channels to the total $\nu_e$ spectrum depends on 
the black hole temperature. One can see from Fig.2, that at high temperature decays of massive particles evaporated by the black hole
become very important at high energy tail of the spectrum (if the corresponding branching ratios are not too small). 

The total instantaneous neutrino spectrum from a black hole evaporation is given by the sum
\begin{equation}
f(E,m)= f_H(E,m)+f^{(\mu)}(E,m)+f^{(q)}(E,m)+ ...\;\;\;,
\end{equation}
and the total background neutrino spectrum is given by the same Eq.(\ref{28}), except the change $f_H(E,m)\to f(E,m)$.

\section{Constraints on the spectral index}
\label{sec:constr}
The spectral index of initial density fluctuations is defined by the relations
\begin{eqnarray}
\label{41}
\sigma_r^2=\frac{1}{V_W^2}\int\frac{d^3k}{(2\pi)^3}\left|\delta_k\right|^2
W_k^2(r)\;\;\;\;,\;\nonumber\\
\\
\left|\delta_k\right|^2=Ak^n\;\;\;.\nonumber
\end{eqnarray}
Here, $\delta_k$ and $W_k$ are Fourier transforms of the density field
$\delta(\vec x)$ and the window function of comoving size $r$, respectively,
$V_r$ is the effective volume filtered by $W_r$.

One  obtains from Eq.(\ref{41}):
\begin{equation}
\sigma_R^2(t)\simeq A\frac{k_{fl}^3}{2\pi^2}\cdot k_{fl}^n\;\;\;,
\end{equation}
where $k_{fl}$ is the comoving wave number, characterizing the perturbed
region, 
\begin{equation}
k_{fl}=\frac{a(t)}{R}\;\;\;,
\end{equation}
and $R$ is the physical size of this region at arbitrary moment $t$.
Using the connection of $k_{fl}$ with the fluctuation mass, 
\begin{equation}
M(t)=\frac{4}{3}\pi\left(\frac{a(t)}{k_{fl}}\right)^3\rho(t)\;\;\;,
\end{equation} 
and introducing the horizon mass $M_h$ (which is equal to the fluctuation mass
at the moment when the fluctuated region crosses horizon) we can rewrite
Eq.(\ref{41}) in the form:
\begin{equation}
\label{45}
\sigma_R(t)=\left(\frac{M_{hor}(t)}{M(t)}\right)^{2/3}\sigma_H(M_h)\;\;\;.
\end{equation}
Here, $M_{hor}$ is the horizon mass at $t$ , $\sigma_H(M_h)$ is the standard
deviation at horizon crossing.

At the initial moment of time one has :
\begin{equation}
M_{hor}(t_i)=M_i\;\;\;,\;\;\;M(t_i)=M\;\;\nonumber.
\end{equation}
The form of the $\sigma_H(M_h)$ function depends on the spectral index:

\begin{equation}
\label{46}
\sigma_H \sim\left\{ 
{
\begin{array}{lcl}
M_h^{\frac{1-n}{4}}\;\;\;&,&\mbox{radiation dominance}\\
M_h^{\frac{1-n}{6}}\;\;\;&,&\mbox{matter dominance}\;\;.
\end{array}}
\right.
\end{equation}
 
Eqs.(\ref{**}) in Sec.\ref{sec:PBHsp} are obtained from Eqs. (\ref{45})-(\ref{46}).

From COBE data we know the normalization of $ \sigma_H(M_h) $ at present horizon
size \cite{18,19}:
\begin{equation}
\sigma_H(M_{h\,,0})=9.5\cdot10^{-5}\;\;\;;\;\;\;M_{h\,,0}=10^{56}g\;\;\;.
\end{equation}
Now one can easily show that $\sigma_H(M_h)$ for radiation dominance case
is connected with $\sigma_H(M_{h\,,0})$ by the following approximate relation:
\begin{equation}
\sigma_H(M_h)=\sigma_H(M_{h\,,0})\left(\frac{M_{h\,,0}}{M_{h\,,eq}}\right)
^{\frac{n-1}{6}}\left(\frac{M_h}{M_{h\,,eq}}\right)^{\frac{1-n}{4}}\;\;\;.
\end{equation}

Our calculation of neutrino spectra from evaporating PBHs contains two 
parameters: a spectral index $n$ and a time of an end of the inflation $t_i$
(which, by assumption, is a time when density fluctuations develop). We 
assume that at $t_i$ the universe has as a result of the reheating the 
equilibrium temperature $T_{RH}$. The connection of $T_{RH}$ and $t_i$
is given by the standard model (in formulas of this section we use 
the convention $\hbar=c=1$):
\begin{equation}
t_i=0.301 g_*^{-1/2}\frac{M_{pl}}{T_{RH}^2}\approx\frac{0.24}{T_{RH}^2(MeV)}
\;\;\;s
\end{equation}
($g_*\sim100$ is the number of the degrees of freedom in the early universe).

\begin{figure}
\caption{Redshift dependence of the of the integrand in the expression (\ref{313}) for a neutrino background spectrum, for two values of the neutrino energy and for several values of the parameter $T_{RH}$.}
\label{fig:fig3}
\epsfig{file=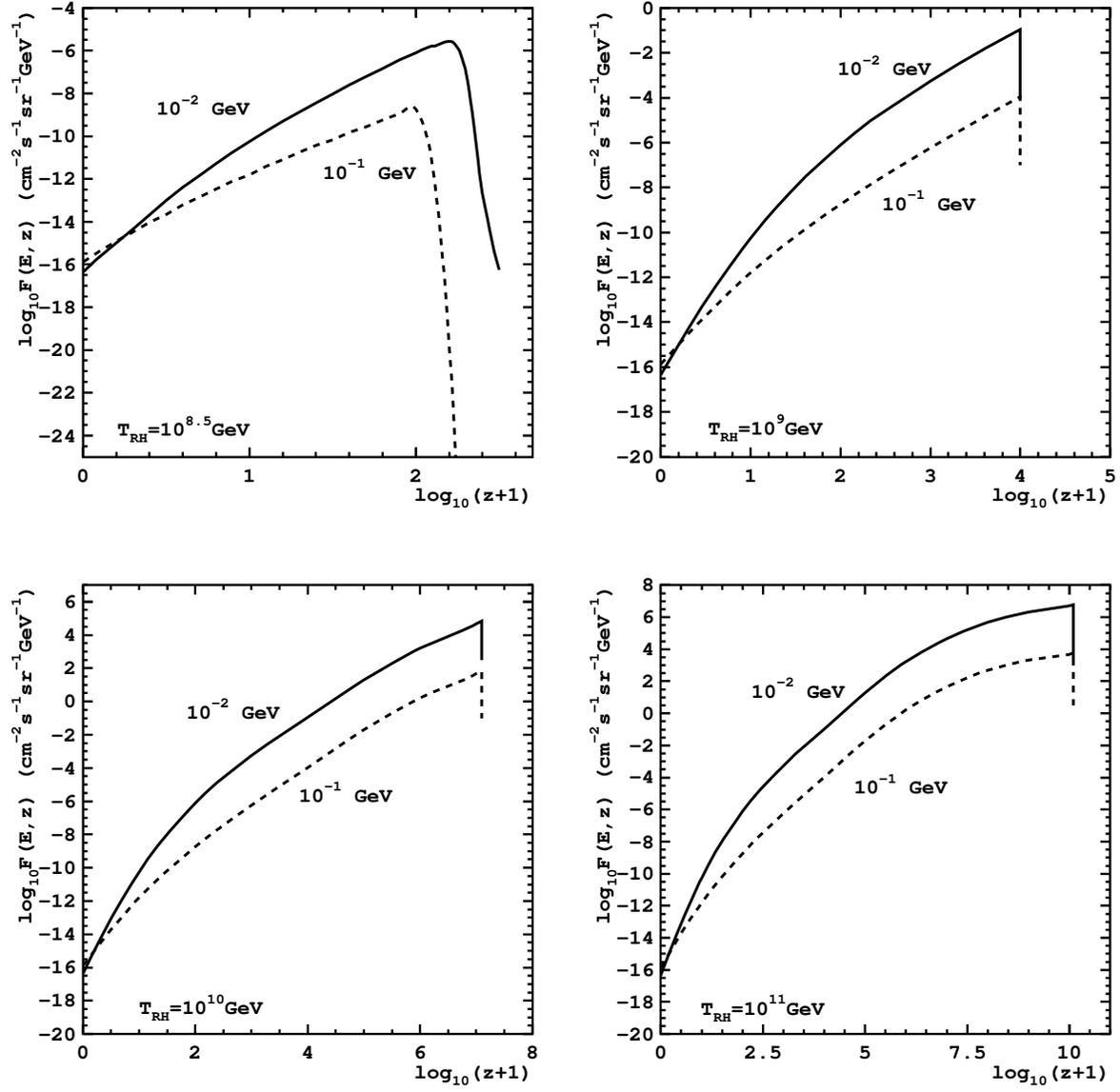,width=\columnwidth}
\end{figure}

On Fig.\ref{fig:fig3} the integrand of the background spectrum integral (Eq.(\ref{313}))
is shown as a function of redshift $z$ for several values of the parameter $T_{RH}$.
 Each 
curve on Fig.\ref{fig:fig3} has a strong cut off on some redshift value. This feature is
connected with the existence in our model the minimum value of PBH mass,
\begin{equation}
M_{BH}^{min}=\gamma^{1/2}M_i\;\;\;.
\end{equation}
The PBH mass spectrum is steeply falling function of the mass, so the masses near minimum give
a largest contribution to the neutrino background. The moment of their evaporation is, approximately,
\begin{equation}
t_{ev}\approx \frac{\left(M_{PBH}^{min}\right)^3}{3\alpha}\;\;,
\end{equation} %
and the corresponding redshift is determined by the relation 
\begin{equation}
z_{ev}+1\approx (z_{eq}+1)\left(\frac{t_{eq}}{t_{ev}}\right)^{1/2}\;\;\;.
\end{equation}
Larger masses evaporate at larger times and smaller redshifts. If, for instance, $T_{RH}=10^{10}\;\;\mbox{GeV}$,
one has $t_i=0.24\cdot 10^{-26}\;\;\mbox{s}$ ; $M_i=7,5\cdot 10^{10}\;\;\mbox{g}$ , $z_{ev}\sim 10^{7}$.
So, at $T_{RH}=10^{10}\;GeV$ the redshifts  of order of $10^7$ give a largest contribution to the neutrino background,
and this is clearly seen at Fig\ref{fig:fig3}.

The cut off value $z_{ev}$ strongly depends on $T_{RH}$: 
\begin{equation}
\label{414}
z_{ev}\sim\frac{1}{\sqrt{t_{ev}}}\sim\left( M_{BH}^{min}\right)^{-3/2}\sim 
t_i^{-3/2}\sim T_{RH}^{3}\;\;.
\end{equation}

\begin{figure}
\caption{Separate contributions to neutrino background spectra for different channels of the neutrino production during PBH evaporation, for fixed values of the parameters $T_{RH}$ and $n$.}
\label{fig:fig4ab}
\epsfig{file=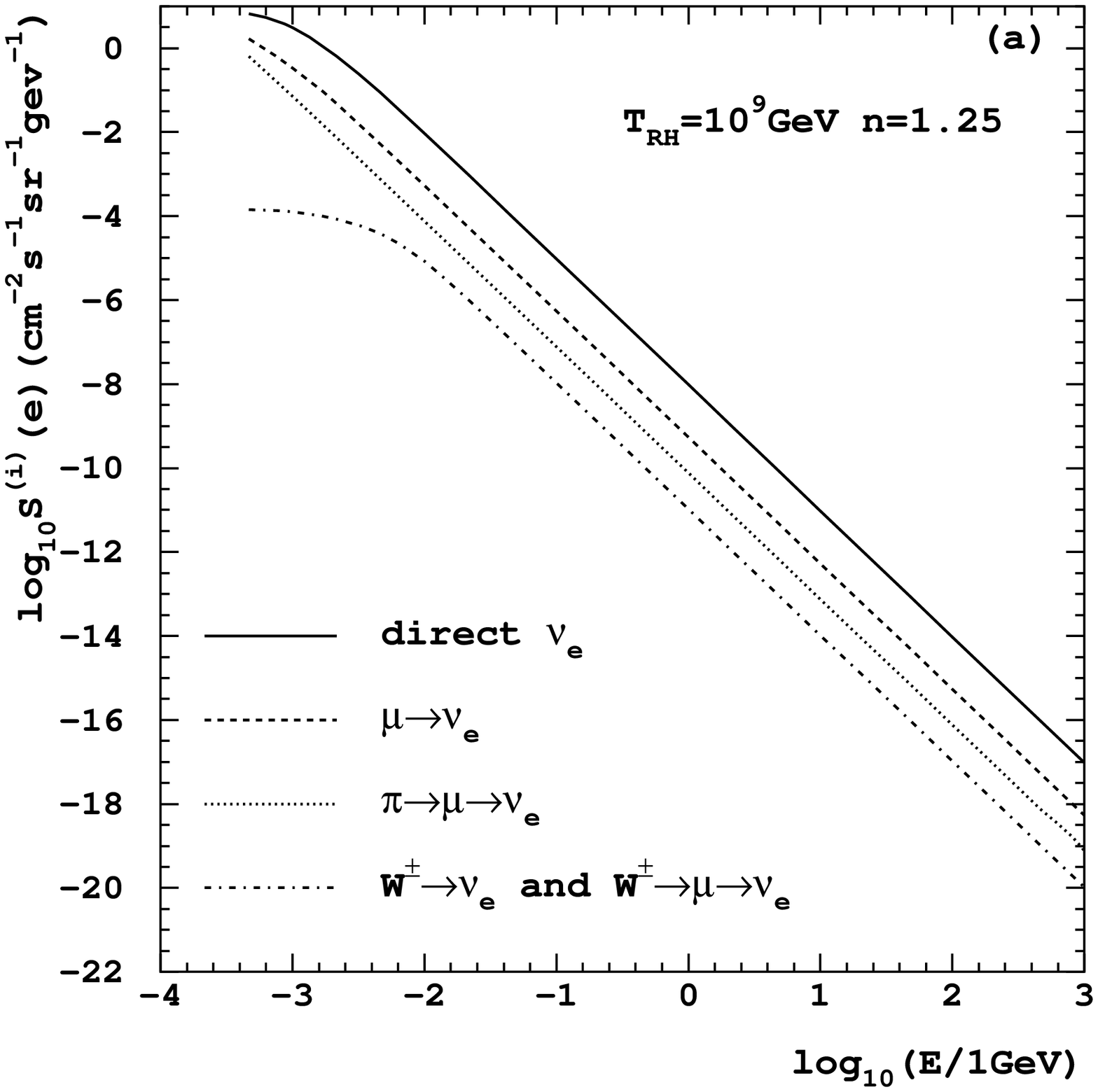,height=11cm}
\epsfig{file=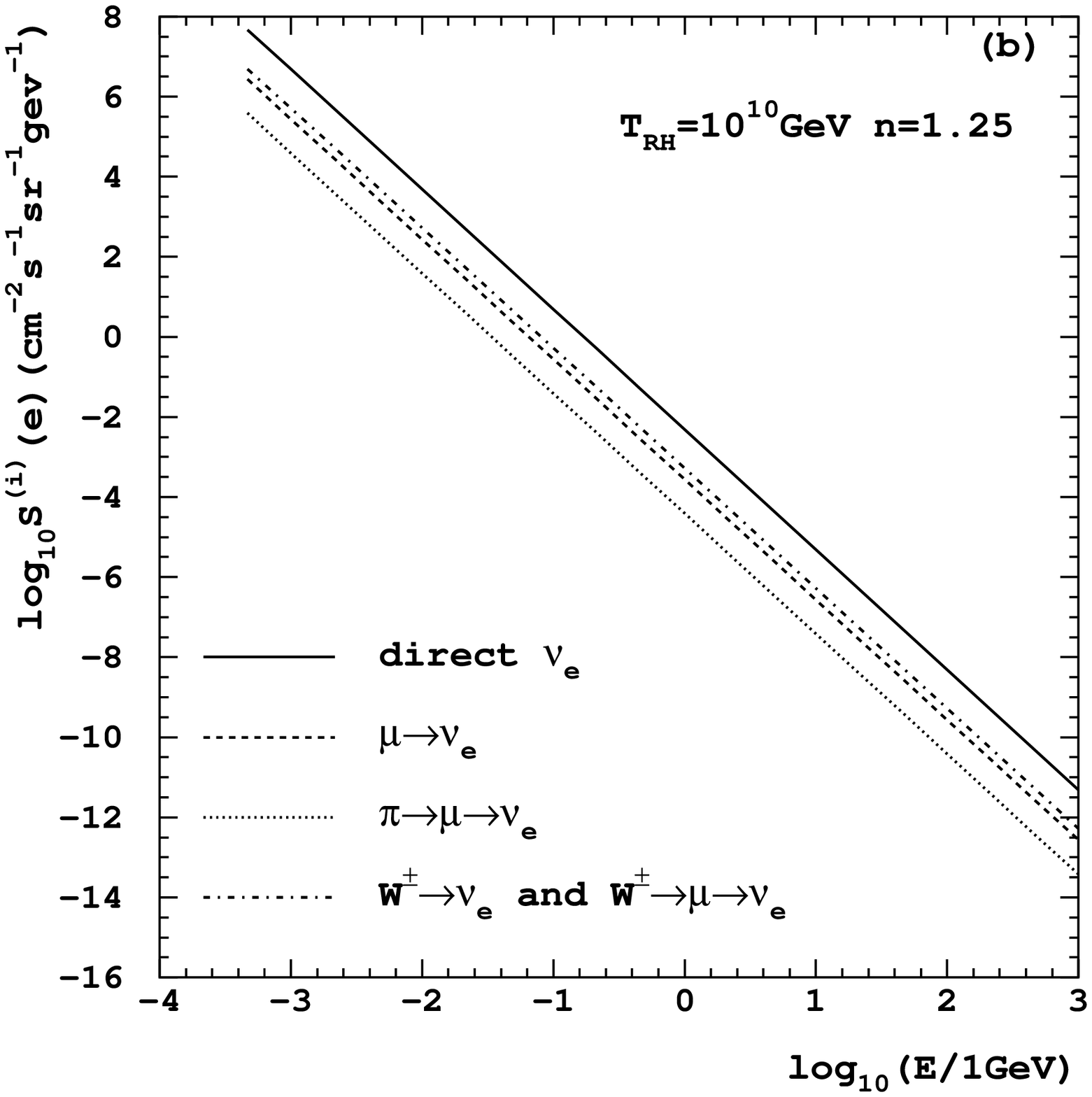,height=11cm}
\end{figure}

On Figs.\ref{fig:fig4ab} spectra of electron neutrino background are shown separately for several channels of the neutrino production.
One can see from these figures that the contribution to the summary background from the
direct $\nu_e$ emission is dominant (at least, for $E\gtrsim 10 \text{ MeV}$) and the relative 
importance of different channels changes with an increase of $T_{RH}$. It is seen also that
the contribution of the quark fragmentation channel is very small at $E\gtrsim 10 \text{ MeV}$
$(\lesssim 1 \% )$, so the evident underestimation of this channel in our calculation, connected with the neglect
of the contribution of heavy quark fragmentations, has no particular importance.
\begin{figure}
\caption{Total energy spectra of electron neutrino background calculated with $n=1.25$ and with different values of the parameter $T_{RH}$.}
\label{fig:fig5}
\epsfig{file=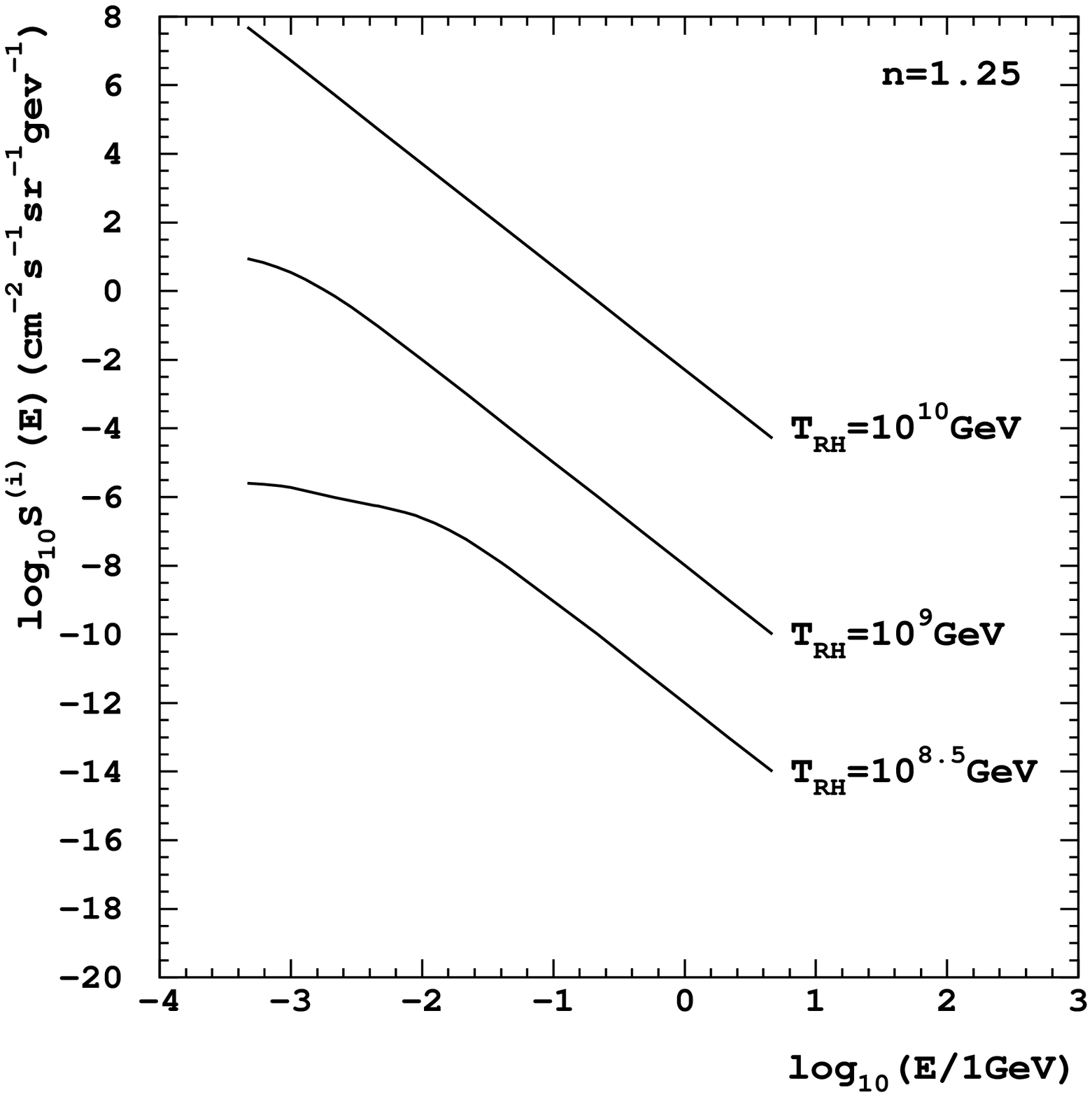,width=\columnwidth}
\end{figure}
\begin{figure}
\caption{Total energy spectra of electron neutrino background calculated with $T_{RH}=10^9\text{ GeV}$ and with different values of the spectral index. The theoretical atmospheric neutrino spectrum at Kamiokande site [] (avereged over all directions) is shown by the dashed line.}
\label{fig:fig6}
\epsfig{file=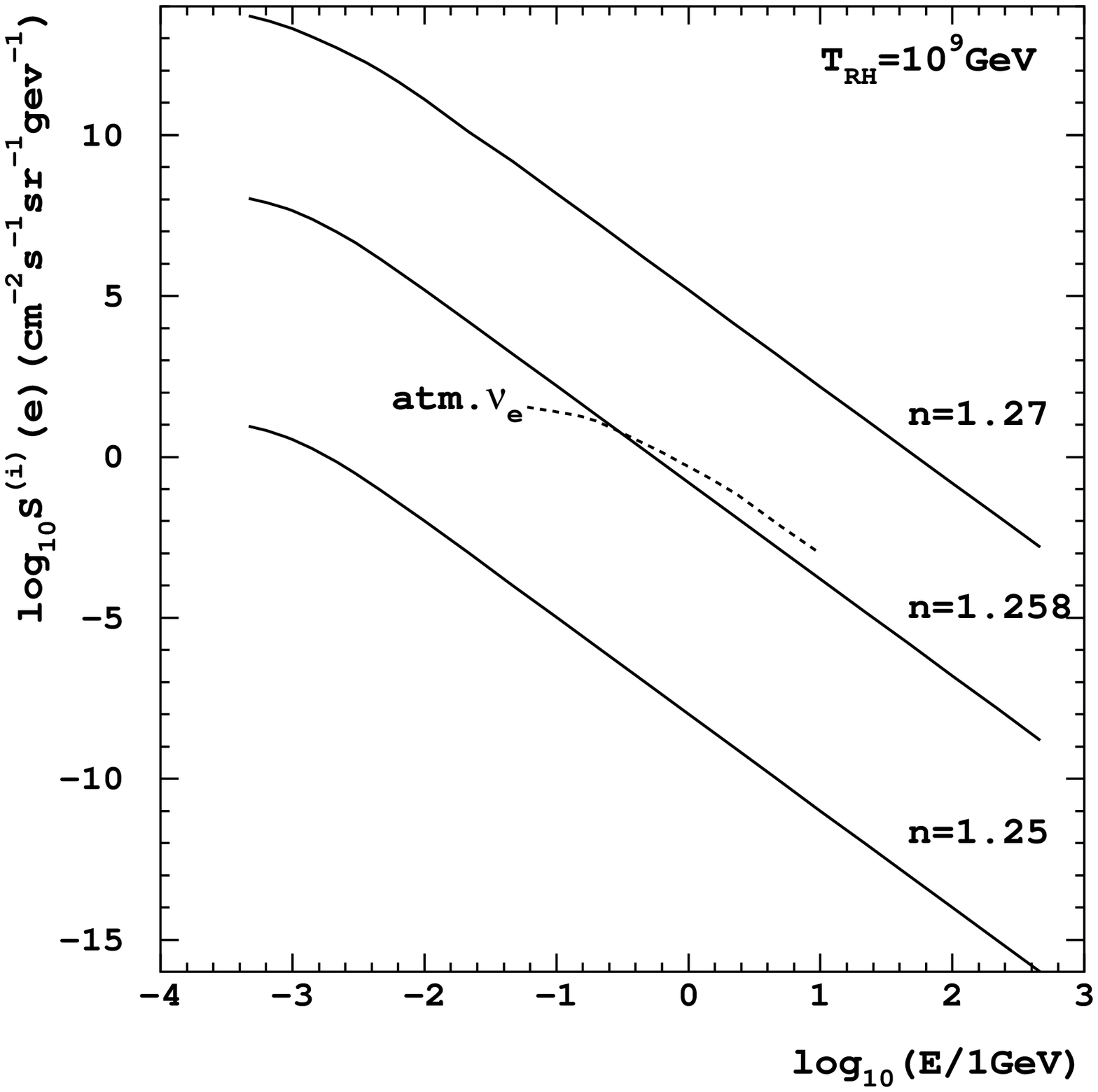,width=\columnwidth}
\end{figure}

Some typical results of electron neutrino background calculations are shown on Figs.\ref{fig:fig5},\ref{fig:fig6}
The main features of such spectra have been revealed in previous works \cite{12,26,27}: $E^{-3}$
dependence above  $100 \text{ MeV}$, the flattening out of the spectra at lower energies and the
shift of the turnover energy with a change of the $T_{RH}$ value \cite{12}.
According to Figs.\ref{fig:fig5},\ref{fig:fig6} the turnover energy is about $1\text{ MeV}$ at $T_{RH}=10^9\; \mbox{\it GeV}$.

For an obtaining of the constraints on the spectral index we use three types
of neutrino experiments.

1. {\it Radiochemical experiments for the detection of solar 
neutrinos.} There are data from the famous Davis experiment \cite{20} and the
$Ga-Ge$ experiment \cite{21}. The cross section for the neutrino absorption via
a bound-bound transition was calculated using the approximate formula
\begin{equation}
\sigma = \left\{
      \begin{array}{lcl}
      \frac{G_F^2}{\pi}\left(\langle 1 \rangle^2+\left(\frac{g_a}{g_v}\right)^2
      \langle\sigma\rangle^2\right)p_e E_e&,&E<100MeV\\
      \mbox{\it const.}&,&E>100MeV\;\;\;.
      \end{array}\right.
\end{equation}
In the case of the Davis experiment ($Cl-Ar$ reaction) we take into account
the super-allowed transition only, for which 
\begin{equation}
\langle 1 \rangle^2=3\;\; ,\;\; \langle\sigma\rangle^2=0.2\;\; ,\;\; E_{thr}\approx 5MeV.\nonumber
\end{equation} 
In the $Ga-Ge$ case the main contribution gives the ground state - ground state
transition ($E_{thr}=0.242MeV$). The corresponding cross section is, 
for $E<100MeV$,   
\begin{equation}
\sigma\left(\nu+Ga\to Ge+e\right)\cong 0.646\cdot 10^{-44}p_e E_e . 
\end{equation}
The number of target atoms $N^T$ is about $2.2\cdot 10^{30}$ for the
$Cl-Ar$ experiment \cite{20} and $\sim 10^{29}$ for the $Ga-Ge$ experiments \cite{21}.
The average statistics is $\sim 1.5\; day^{-1}$ ($Cl-Ar$) and $\sim 1\; 
day^{-1}$ ($Ga-Ge$). The constraint is calculated using the relation
\begin{equation}
4\pi\cdot N^T\cdot10^5\cdot\int S(E)\sigma(E)dE<1.
\end{equation}

2. {\it The experiment on a search of an antineutrino flux from the Sun} \cite{22}.
In some theoretical schemes (e.g., in the model of a spin - flavor precession in
a magnetic field) the Sun can emit rather large flux of antineutrinos. LSD
experiment \cite{22} sets the upper limit on this flux, $\Phi_{\tilde \nu}/\Phi_{\nu}
\le 1.7\%$. In this experiment the neutrino detection is carried out using
the reaction
\begin{equation}
\label{reaction}
\tilde \nu _e + p \to n + e^+
\;\;.
\end{equation}
The number of target protons is $\sim 8.6\cdot 10^{28}$ per 1 ton of the
scintillation detector,  and the obtained upper limit is $0.28$ antineutrino
events per year per ton \cite{22}. The cross section of the reaction~(\ref{reaction})
is well known (see, e.g.,\cite{23}).
It grows with the neutrino energy up to $E\sim 2\;GeV$, and is a constant ($\sim 0.5\cdot 10^{-38}\;cm^{2}$)
at larger energies.The product of this cross section and a PBH antineutrino background spectrum has the maximum
at $E\sim 100MeV$. The constraint is determined from the condition that the calculated effect in the {\bf LSD}
detector is smaller than the upper limit obtained in ref.\cite{22}

3. {\it The Kamiokande experiment on a detection of atmospheric electron 
neutrinos }\cite{24}. In this experiment the electrons arising in the reaction  
\begin{equation}
\nu_e^{atm}+n\to p+e^-
\end{equation}
in the large water Cherenkov detector were detected and, moreover, their 
energy spectrum was measured. This spectrum has a maximum at the energy
about $300MeV$. The spectrum of the atmospheric electron neutrinos is
calculated(see, e.g.,\cite{25}) with a very large accuracy (assuming an absence of the neutrino 
oscillations) and the experimentally measured electron spectrum coincides,
more or less, with the theoretical prediction. The observed electron excess
at $E\sim 100MeV$ (which is a possible consequence of the oscillations) 
is not too large. We use the following condition for an obtaining  the our 
constraint : the absolute differential intensity of the PBH neutrino 
background at the neutrino energy $E\sim 0.3GeV$ cannot exceed the theoretical
differential intensity of the atmospheric electron neutrinos at the same 
energy (otherwise the total electron energy spectrum is strongly different from
the observed one).

\section{Results and discussions}

\label{sec:res}

\begin{figure}
\caption{Constraints on the spectral index $n$ as a function of the reheating temperature $T_{RH}$ from three types of the neutrino experiments.}
\epsfig{file=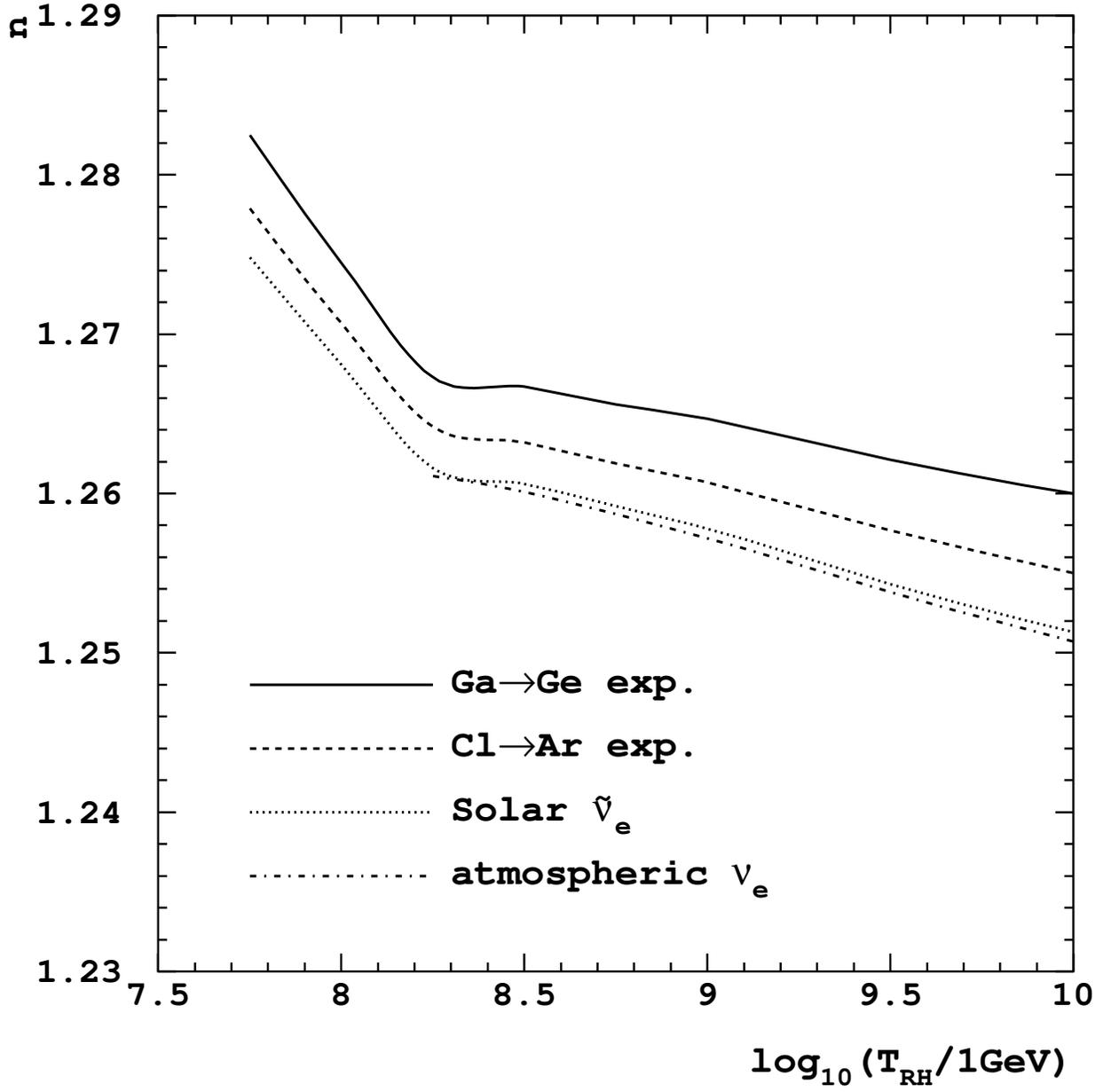,width=\columnwidth}
\label{fig:fig7}
\end{figure}

Fig.\ref{fig:fig7} shows our results for the spectral index constraints. It is seen that 
the best constraints are obtained using the Kamiokande atmospheric neutrino data and
the LSD upper limit on an antineutrino flux from the Sun. 

The behavior of the 
constraint curve $n(T_{RH})$ is sharply different from that was obtained 
in ref. \cite{13} (authors of ref. \cite{13} used the initial PBH mass function following
from the near critical collapse scenario with a domination of the earliest 
epoch of PBHs formation and diffuse extragalactic photon data).

The slightly non-monotonic behavior of the curves Fig.\ref{fig:fig7} near $T_{RH}\sim 3\cdot 10^8\;\mbox{GeV}$
is connected with the sharp increase of the function $\alpha(m)$ at $m\approx 10^{14}\:g$ due to a switching
on the  quark evaporations (see Fig.\ref{fig:fig1}).
Indeed, according to the relations Eq.(\ref{414}), the minimum value of PBH mass in the initial mass spectrum is inversely proportional
to $T_{RH}^2$. More exactly, the relation between $M_{BH}^{min}$ and $T_{RH}$ is 
\begin{equation}
M_{BH}^{min}(g)=\frac{7.2\cdot 10^{30}}{T_{RH}^2(GeV)}\;\;. 
\end{equation}
Thus, when the parameter $T_{RH}$ passes over the value $\sim 3 \cdot 10^8 \;GeV$, $M_{BH}^{min}$ becomes smaller than $10^{14}\;g$,
and the value of $\alpha$ used in Eqs.(\ref{311}-\ref{28}) increases on $\sim 20$ units.

One can compare our spectral index constraints with the corresponding results
of ref. \cite{12}, where the same initial mass spectrum of PBHs had been used. Some
difference in resulting constraints ($\Delta n\sim 0.02$) may be connected simply 
with the fact that authors of ref. \cite{12} use slightly different formula for
$\sigma_H (M_h)$, namely,
\begin{equation}
\sigma_{H}(M_h)=\sigma_H(M_{h\;,0})\left(\frac{M_h}{M_{h\;,0}}\right)^{\frac
{1-n}{4}}.
\end{equation}
In other respects the constraints are quite similar although in ref. \cite{12} 
they were based on diffuse extragalactic photon background data. 

Constraints on the spectral index, obtained using the neutrino background calculations, are sensitive
to a possible existence of a {\bf QCD} photosphere \cite{28,29} around black holes only in one case:
if the density of the outward-propagating plasma is so high that it is opaque for
neutrinos. It was shown above (see fig\ref{fig:fig4ab}) that the fragmentations of evaporated quarks 
are inessential and can be neglected when calculating the electron neutrino background (in spite of the fact
that neutrinos from these fragmentations are very abundant (fig.\ref{fig:fig2})). The reason is simple:
neutrino energy is strongly redshifted during cosmological expansion. It is clear that the fragmentations
of photosphere's quarks are also inessential especially as energies of these quarks are, on average, lower
than energies of quarks evaporated directly. Thus, only the neutrino opacity of the photosphere can modify
our results. According to a general phylosophy of a photosphere's model an opaque "neutrino photosphere"
eventually forms \cite{28}, but it may happen too late, when the black hole mass is already too small
to give noticeable contribution to the integral in Eq.(\ref{28}).

A precise calculation of the PBH neutrino background must include also a taking into account of the neutrino absorption during a travelling in the space.
In the case of the photon background the most important absorption process is a pair production on neutral matter \cite{14} due to which the photons from 
PBHs evaporated earlier than $z\approx 700$ are absent today. In our case, the analog of an optical depth of the universe for the neutrino emitted at
a redshift z and having today an energy $E$ is given by the integral
\begin{equation}
\label{RRR}
\tau (z,E) = c \int\limits_0^z\sigma\left( E(1+z')\right)\cdot n(z') \frac{d t}{d z'} d z'\;\;.
\end{equation}
Here, $\sigma (E)$ is the neutrino interaction cross section, $n(z)$ is a number density of the target particles. 

Two processes are potentially
"dangerous": neutrino-nucleon inelastic scattering growing linearly with an energy, and annihilations with neutrinos of the relic background:
\begin{eqnarray}
\nu_e + N \to e^{-} + anything,\nonumber\\
\\
\nu_e + \tilde\nu_e (relic) \to \sum\limits_i (f_i +\tilde f_i).
\end{eqnarray}
Here, $f_i$ are charged fermions (leptons and quarks). As is known, the relic neutrino background exists, with a Planck distribution,
from the epoch of neutrino decoupling ($z\approx 10^{10}$).

For estimations of the neutrino absorption it is enough to use some characteristic value of a neutrino energy. For the neutrino processes in the detectors,
discussed in the previous Section, such value is about $100 MeV$. In the case of the $\nu N$-scattering one has, approximately,
\begin{equation}
\sigma_{\nu N}(E) \sim 10^{-38}\cdot \left(\frac{E}{\mbox{\it GeV}} \right) cm^{-2}\;\;,
\end{equation}
\begin{equation}
\frac{n_N(z)}{(1+z)^3}= n_{0\!N}\sim (10^{-6}-10^{-7})\;cm^{-3}\;\;.
\end{equation}
Using these values and the formula (\ref{311}) for ${d t}/{d z}$, one can easily estimate the integral (\ref{RRR}) for $\tau (z,E)$. One obtains
the important result:
\begin{equation}
\tau_{\nu N}(z\approx10^7,E=100MeV)\ll 1\;\;.
\end{equation}

In the case of the absorption through annihilations of evaporated neutrinos with relic neutrinos the integral (\ref{RRR}) reduces to \cite{30}
\begin{equation}
\label{DOUBLEASTERIX}
\tau_{\nu \tilde \nu }(z,E) \simeq 0.3 \cdot 10^{-34}\left(\frac{E}{\mbox{\it TeV}} \right)\cdot\frac{1}{H_0}\int\limits_0^z (1+z')^5 \frac{d t}{d z'}d z'\;\;.
\end{equation}
This  gives
\begin{equation}
\tau_{\nu \tilde\nu}(z\simeq 2\cdot 10^6,E=100 \mbox{\it MeV})\le 1\;\;.
\end{equation}

From here one can conclude that the neutrino absorption becomes important (for the problems considered in this paper) at $z\sim 2\cdot 10^6$.
It was shown in Sec.\ref{sec:constr} that at $T_{RH} = 10 ^{10}\;GeV$ the redshifts $\sim 10^7$ give a maximum contribution to the neutrino 
background (see Fig.\ref{fig:fig3}). Besides, it was shown that, in general, a redshift value corresponding to such a maximum is proportional
to $T_{RH}^3$ (Eq.(\ref{414})). So, the value $z=2\cdot 10^6$ corresponds to $T_{RH}\approx 6\cdot 10^9 \;GeV$, and, therefore, for $T_{RH}> 
6\cdot 10^9\; GeV$ the neutrino absorption effects are important.

One should note, at the end, that 
usually the calculations of these constraints are accompanied by the calculation
of the bounds based on requirement that the energy density in PBHs does not overclose
the universe at any epoch ($\Omega_{BH} < 1$). For a setting of such bounds one 
must consider the cosmological evolution of the system PBHs + radiation. We 
intend
 to carry out these calculations in a separate paper.

\acknowledgments

We wish to thank G. V. Domogatsky for valuable discussions and comments.
 
We are grateful also 
to H.I.Kim for informing us about his work on the same problem and for useful remarks.

\end{document}